\documentclass[lettersize,journal]{IEEEtran}

\usepackage{amsmath,amsfonts}
\usepackage{algorithmic}
\usepackage{algorithm}
\usepackage{array}
\usepackage[caption=false,font=normalsize,labelfont=sf,textfont=sf]{subfig}
\usepackage{textcomp}
\usepackage{stfloats}
\usepackage{url}
\usepackage{verbatim}
\usepackage{graphicx}
\usepackage{color}
\usepackage{cite}
\usepackage{xcolor}
\usepackage{cancel}

\newcommand\red[1]{\textcolor{red}{#1}}
\newcommand\blue[1]{\textcolor{blue}{#1}}
\newcommand\green[1]{\textcolor{green}{#1}}
\newcommand\magenta[1]{\textcolor{magenta}{#1}}
\usepackage[normalem]{ulem}
\usepackage{ifthen}
\usepackage{xspace}
\newcounter{showedits}
\newcounter{forreferee}
\newcommand\rsout[1]{\ifthenelse{\value{showedits}=0}{}{\red{\sout{#1}}}}
\newcommand\gsout[1]{\ifthenelse{\value{forreferee}=0}{\ifthenelse{\value{showedits}=0}{}{\green{\sout{#1}}}}{\rsout{#1}}}
\newcommand\gsoutmath[1]{\ifthenelse{\value{forreferee}=0}{\ifthenelse{\value{showedits}=1}{}{\green{\cancel{#1}}}}{\ifthenelse{\value{showedits}=0}{}{\red{\cancel{#1}}}}}
\newcommand\brep[1]{\ifthenelse{\value{showedits}=0}{#1}{\blue{#1}}}
\newcommand\bsout[1]{\ifthenelse{\value{forreferee}=0}{\ifthenelse{\value{showedits}=0}{}{\blue{\sout{#1}}}}{}}
\newcommand\mrep[1]{\ifthenelse{\value{forreferee}=0}{\ifthenelse{\value{showedits}=0}{#1}{\magenta{#1}}}{\brep{#1}}}

\usepackage{hyperref} 

\begin{document}

\title{Low-Frequency Noise Performance of Microstrip-Coupled Lumped-Element Aluminum KIDs using Hydrogenated Amorphous Silicon Parallel-Plate Capacitors for NEW-MUSIC}

\author{%
\IEEEauthorblockN{%
Simon Hempel-Costello\IEEEauthorrefmark{1},
Sunil R. Golwala\IEEEauthorrefmark{1},
Andrew D. Beyer\IEEEauthorrefmark{2},
Daniel Cunnane\IEEEauthorrefmark{2},
Peter K. Day\IEEEauthorrefmark{2},
Fabien Defrance\IEEEauthorrefmark{2},
Clifford F. Frez\IEEEauthorrefmark{2},
Adriana Gavidia\IEEEauthorrefmark{1},
Junhan Kim\IEEEauthorrefmark{3},
Jean-Marc Martin\IEEEauthorrefmark{1},
Yann Sadou\IEEEauthorrefmark{1},
Jack Sayers\IEEEauthorrefmark{1},
Shibo Shu\IEEEauthorrefmark{4},
Shiling Yu\IEEEauthorrefmark{1}\IEEEauthorrefmark{5}\IEEEauthorrefmark{6}%
} \\[0.5em]
\IEEEauthorblockA{\IEEEauthorrefmark{1}Division of Physics, Mathematics, and Astronomy, California Institute of Technology, Pasadena, CA, USA, 91125} \\
\IEEEauthorblockA{\IEEEauthorrefmark{2}Jet Propulsion Laboratory, California Institute of Technology, Pasadena, CA, USA, 91109} \\
\IEEEauthorblockA{\IEEEauthorrefmark{3} Korea Advanced Institute of Science and Technology, Daejeon, Korea, 34141}
\IEEEauthorblockA{\IEEEauthorrefmark{4}Institute of High Energy Physics, Chinese Academy of Sciences, Beijing, China, 100049} \\
\IEEEauthorblockA{\IEEEauthorrefmark{5}National Astronomical Observatories of China, Beijing, China, 100101} \\
\IEEEauthorblockA{\IEEEauthorrefmark{6}University of the Chinese Academy of Sciences, Beijing, China, 101408} \\
\thanks{Manuscript received August 8, 2025; revised August 16, 2025.}%
}

\markboth{IEEE Transactions on Applied Superconductivity}%
{Hempel-Costello \MakeLowercase{\textit{et al.}}: Low-Frequency Noise Performance of Al/a-Si:H MS-PPC-LEKIDs for NEW-MUSIC}


\maketitle
\begin{abstract}
We present measurements of the low-frequency noise 
of 
\mrep{a} microstrip-coupled, lumped-element 
aluminum kinetic inductance \gsout{detectors}\mrep{detector design} 
that use\mrep{s a} hydrogenated amorphous silicon parallel-plate capacitor\gsout{s} (Al/a-Si:H MS-PPC-LEKID\gsout{s}), \gsout{which are}\mrep{a design} under development 
for the Next-generation Extended Wavelength MUltiband Submillimeter Inductance Camera (NEW-MUSIC)\mrep{\ on the Leighton Chajnantor Telescope (LCT)}. We show that, under dark conditions, \rsout{these devices are }\mrep{this design's noise PSD is dominated by }generation\mrep{-}recombination (GR) noise \rsout{dominated }down to 0.1 Hz\mrep{.}\rsout{ and, under
optical load, they are likely dominated by GR and photon noise
down to tenths of a Hz and possibly lower, both in spite of the
use of a-Si:H PPCs.}
Our measurements set \mrep{upper} limits on the low-frequency \mrep{(0.01--1~Hz) }two-level-system (TLS) noise of the a-Si:H material that are consistent with \gsout{higher frequency}\mrep{prior} measurements \gsout{in the}\mrep{at} 0.1--10~kHz\gsout{ regime}.
\mrep{By extrapolating these measurements to optically loaded operating conditions and comparing to prior measurements above 100~Hz of noise dominated by photon and quasiparticle statistics under optical load, we project that the Al/a-Si:H MS-PPC-LEKID design's noise PSD will be photon- and generation-recombination-noise-dominated down to tenths of Hz under expected telescope on-sky optical loads.}
These results \gsout{\rsout{establish}\bsout{\brep{suggest}} that our MS-PPC-LEKID design for NEW-MUSIC will be photon-noise-limited under a range of observing conditions and, more generally, that a-Si:H PPC-KIDs}\mrep{are an important milestone toward demonstrating that Al/a-Si:H MS-PPC-LEKIDs, and a-Si:H PPC-KIDs more generally,} are a viable new detector technology for even low modulation-rate applications such as astronomy.
\end{abstract}


\section{Introduction}

\IEEEPARstart{T}{he} Next-generation Extended Wavelength MUltiband Submillimeter Inductance Camera (NEW-MUSIC~\cite{SPIE}) is an imaging polarimeter with six co-pointed spectral bands spanning 80–420 GHz 
under development for 
the Leighton Chajnantor Telescope (LCT). 
%
This set of bands will provide access to \rsout{a }a wide range of scientific targets\rsout{,} including: \gsout{high-frequency synchrotron radiation from }
a variety of transient and time-domain astronomical sources such as \gsout{active galactic nuclei and supernovae, gamma-ray bursts, and tidal disruption events expanding in a dense circumstellar medium}\mrep{explosive or disruptive stellar death in a dense circumstellar medium, outbursts from and accretion onto the remnants, growing young stars, flaring stars, and supermassive black holes via high-frequency synchrotron or cyclotron emission or emission from thermal dust};
diffuse gas 
in galaxy clusters and the circumgalactic medium 
via the Sunyaev–Zeldovich effect; 
dusty, polarized sources in the Milky Way and nearby galaxies; and dusty, star-forming galaxies.

To observe simultaneously across such a large bandwidth over a compact field-of-view, 
NEW-MUSIC employs frequency-selective hierarchical summing of superconducting phased-array slot-dipole antennas
~\cite{newmusic_ltd20_martin, SPIE, newmusic_ltd21_huang},
which enables 
the pixel size to scale with wavelength across 
80--420 GHz. 
These antennas output light onto \gsout{the }superconducting microstripline, necessitating a microstripline-coupled detector.  
Simultaneously, the detectors must be compact and easily multiplexed because each 6.66~mm side-length pixel feeds 42 individual detectors.  Kinetic inductance detectors (KIDs) offer excellent multiplexability, but conventional KID designs \rsout{using interdigitated capacitors (IDCs) }are not sufficiently compact\mrep{\ for this application}, are not easily microstripline-coupled, and can suffer from direct absorption of light in the inductor and/or the capac\mrep{i}tor\gsout{s}\brep{\mbox{\cite{SPIE-2012}}}.

\protect\bsout{The large detector footprints of conventional KIDs arise from challenges in forming multilayer structures with crystalline dielectrics, which necessitate the use of interdigitated capacitors (IDCs)\mbox{\cite{defrance}}.}
\mrep{Conventional KID designs have large footprints because they use single-metal-layer capacitors --- co-planar waveguide (CPW) or interdigitated capacitor (IDC) structures.  Parallel-plate capacitors (PPCs), an inherently compact alternative, have not previously been practical\rsout{: d}\brep{. D}epositable amorphous dielectrics typically have too much two-level-system (TLS) noise, and PPCs using lower-noise crystalline dielectric require either backside etching~\cite{siddiqi2011} --- which can be incompatible with backside optical illumination --- or challenging fabrication processes~\cite{denis_ltd13, beyer_wolte2016}.  Microstripline coupling can be difficult for conventional KID designs because the KID inductor must present an impedance-matched termination to the microstripline.}

\begin{figure}[t]
\centering
\includegraphics[width=0.5 \textwidth]{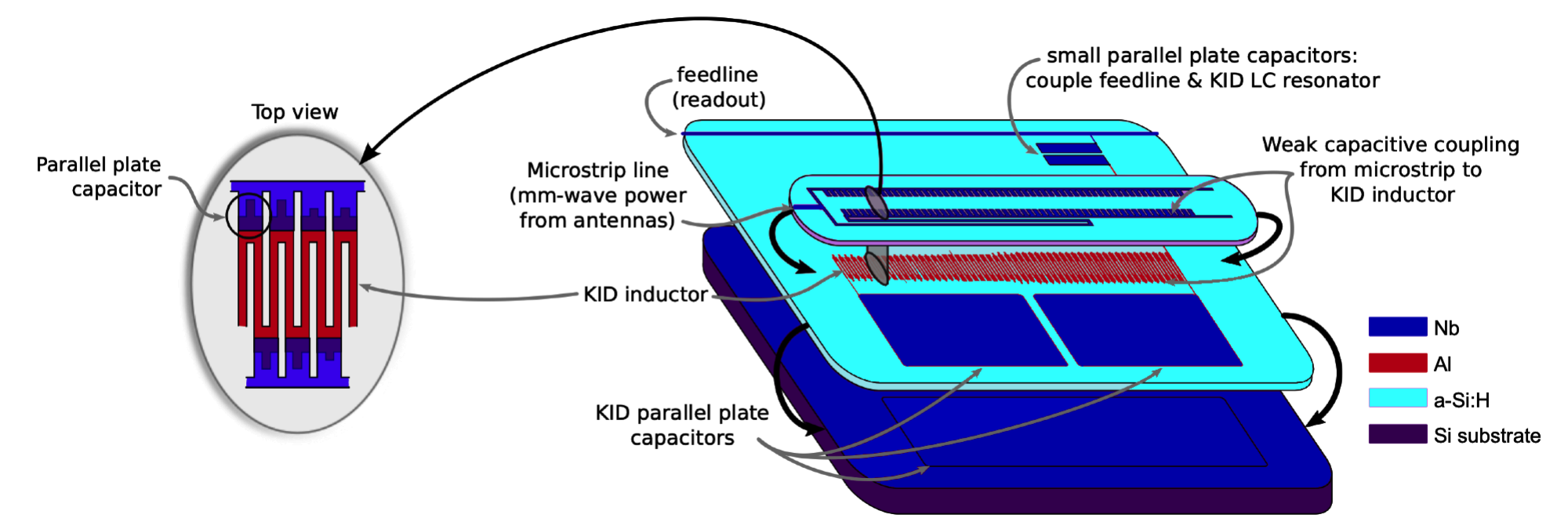}
\caption{\protect\bsout{Design of the PPC-LEKIDs used in NEW-MUSIC. Use of a parallel plate capacitor (PPC) is enabled by the a-SI:H dielectric, which lowers TLS noise dramatically\mbox{\cite{SPIE}}.}\protect\mrep{Aluminum/hydrogenated amorphous silicon, microstrip-coupled, parallel-plate capacitor, lumped-element KID (Al/a-Si:H MS-PPC-LEKID) design.  The PPC consists of two PPCs in series with the Nb ground plane as the center plate for ease of fabrication and maximal mitigation of direct absorption.  The center plate is a virtual ground because of the mirror symmetry of the inductor and the PPC, so there is no cutout in the ground plane for the capacitor pad.  The inset highlights the capacitive coupling of light from the microstripline structure to the inductor.  A more detailed description of the design can be found in\mbox{\cite{SPIE}}.}}
\label{fig:kid_design}
\end{figure}

To address \gsout{this need}\mrep{these challenges}, we have been developing \gsout{Al/a-Si:H }\mrep{a }microstripline-coupled, parallel-plate capacitor, lumped-element KID\gsout{s} (MS-PPC-LEKID\gsout{s}) \mrep{design using aluminum and hydrogenated amorphous silicon (a-Si:H)}~\cite{newmusic_spie2014_ji, newmusic_ltd19_shu, newmusic_ltd20_martin, SPIE}, \brep{a schematic of which} \bsout{are included in FIG.}\mrep{is shown in Fig.~}\ref{fig:kid_design}. In this unique KID architecture, a microstripline capacitively couples millimeter/submillimeter waves to an Al KID inductor that sits above a Nb ground plane, providing shielding against direct absorption.  The KID incorporates a \gsout{parallel-plate capacitor}PPC\gsout{) that is both }\mrep{. PPCs are inherently }compact\mrep{.  Our implementation integrates the PPC with a ground plane so it }\gsout{and }has inherently low susceptibility to direct absorption\mrep{\ (see\mbox{\cite{SPIE}} for details)}.\bsout{This unique KID architecture incorporates a parallel-plate capacitor (PPC) that is both compact and has inherently low susceptibility to direct absorption. Moreover, the microstripline capacitively couples millimeter/submillimeter waves to an Al KID inductor that sits above a Nb ground plane, providing shielding against direct absorption.}

This architecture requires a dielectric film that has low THz loss for the microstripline and low GHz loss and noise for incorporation into the KID PPC.  
\cite{defrance-c, defrance, SRON-Hahnle,SRON-Buijtendorp} have shown that deposited hydrogenated amorphous silicon (a-Si:H) dielectric can 
provide loss and TLS noise performance for GHz resonators comparable to 
that of IDCs, but TLS noise measurements to date have only been in the 0.1-10~kHz modulation frequency regime and have been done using Nb superconducting resonators, leaving un\gsout{proven}\mrep{tested} the performance of a-Si:H PPCs for KID astronomical detectors.

In this paper, we characterize the low-frequency noise performance of \mrep{this }Al/a-Si:H MS-PPC-LEKID\mrep{\ design}\gsout{s, }\rsout{showing they are generation-recombination- or photon-noise limited under dark and optical operation, respectively, down to 0.1 Hz}\brep{.  These measurements demonstrate this KID architecture is stable down to modulation frequencies relevant for astronomical observations
(\bsout{for example}\mrep{\textit{e.g.},} corresponding to 5$^\circ$ on the sky at a scan speed of 0.5$^\circ$/s), set limits on the TLS noise of our a-Si:H material} \bsout{to much lower frequency than has been explored before.}\mrep{at lower frequencies than previously explored, and can be extrapolated to optical operating conditions to project that Al/a-Si:H MS-PPC-LEKIDs will have sensitivity limited only by photon and quasiparticle statistics down to tenths of a Hz.}

\section{Two-Tone Noise Measurement Setup}
\label{sec:two-tone}
Motivated by the desire to make measurements well below 1~Hz, 
we developed a two-tone IQ measurement system
, which enables the use of a monitor tone to measure and remove multiplicative low-frequency noise frequently seen in RF amplifiers and readout systems.
\mrep{While we were not able to complete the commissioning of the correlated noise removal for this work and thus present only ``uncleaned'' noise measurements, we describe it here because we used it, rather than a standard single-tone IQ mixer setup, for the low-frequency dark measurements presented below.  Also, even without cleaning, the setup mitigates the low-frequency additive noise of the final amplifier usually present in such setups.  This brief description also serves to disseminate the concept, which combines the correlated noise removal capability of multi-tone readout systems with the simplicity of a single-tone IQ mixer setup.}

\begin{figure}[!t]
\centering
\includegraphics[width=0.45 \textwidth]{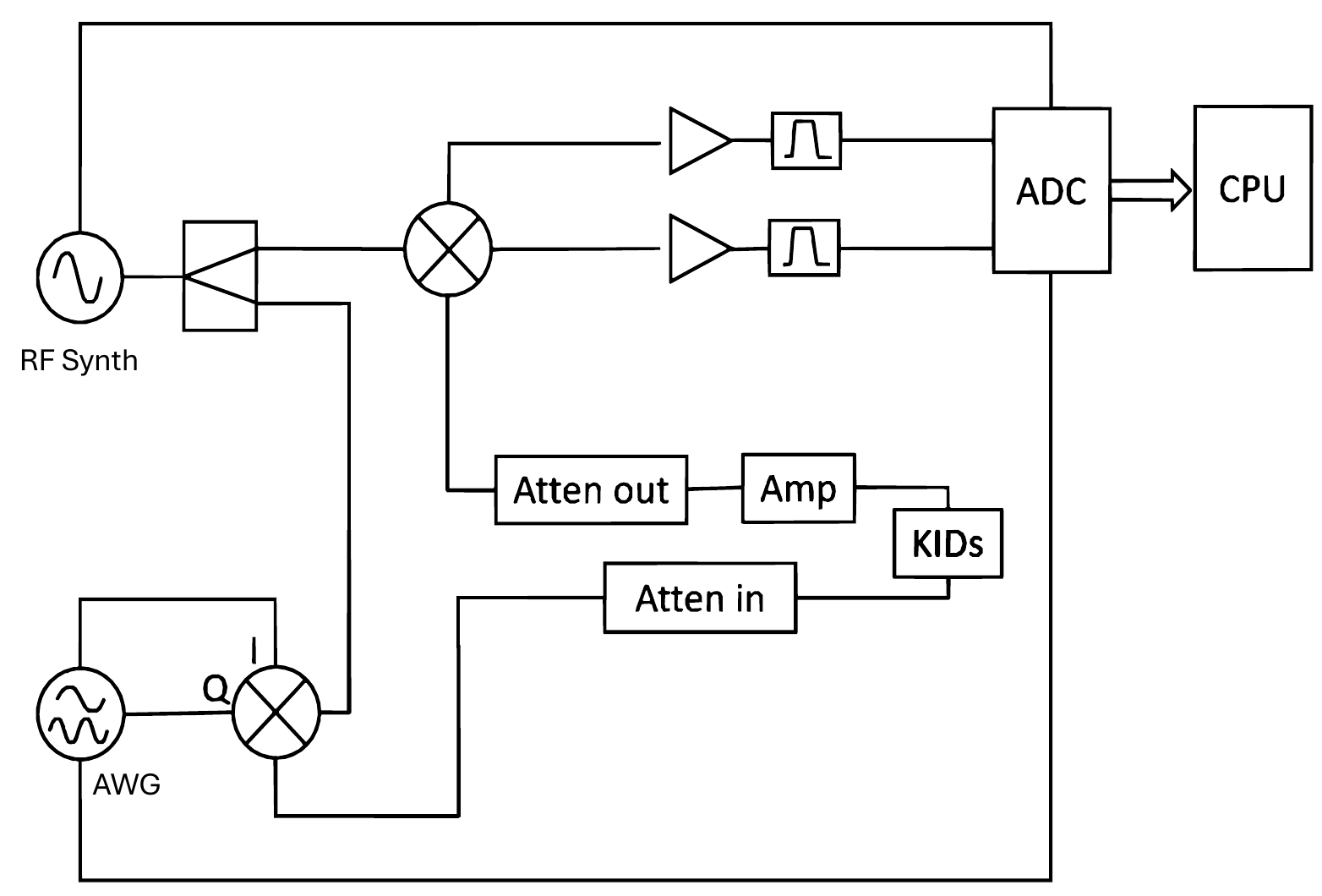}
\caption{Two-tone IQ measurement setup. 
I and Q signals at tens to hundreds of kHz produced by the arbitrary waveform generator (AWG)
are up-mixed with a local oscillator (LO) signal from a RF synthesizer.
The two-tone signal is sent into the cryostat to probe the \protect\rsout{KID on and off resonance}\protect\mrep{device under test}.  
The output signal is down-mixed with a copy of the original LO, amplified and anti-alias filtered at baseband, and digitized.  Final demodulation of the two tones is done in software.
The scheme enables simultaneous acquisition of on- and off-resonance tones 
to enable removal of correlated multiplicative electronics noise. 
It also mitigates baseband amplifier additive low-frequency noise.}
\label{fig:Two Tone Setup}
\end{figure}

The setup, shown in Fig.~\ref{fig:Two Tone Setup}, uses IQ mixing of 
a two-tone baseband signal (provided by an arbitrary waveform generator)
with a local oscillator (LO)  
to 
produce 
on- and off-resonance probe tones. 
The up-mixed tones are applied to the \rsout{KID}\mrep{device under test}, amplified cryogenically and at room temperature, IQ-mixed back to \gsout{their original generation }\mrep{the base}band, amplified again, anti-alias filtered, and digitized.  
Final demodulation and analysis are performed numerically.  In addition to providing a monitor tone, this setup also circumvents the low-frequency additive noise of the generation-band amplifiers by placing both tones at tens to hundreds of kHz.  The system is functionally equivalent to a software-defined-radio system with only two tones and a few hundred kHz baseband bandwidth.  It has the advantage that all the components are easily accessible and individually testable.
\rsout{
While we were not able to complete the commissioning of the off-resonance tone for correlated noise removal for this work and thus present only ``uncleaned'' noise measurements, the setup inherently provides excellent mitigation of additive 1/f noise from the baseband amplifiers for the on-resonance tone alone.
}
\section{TLS Noise Expectation}
\label{sec:tls}

\begin{figure}[t]
\centering 

\includegraphics[width = 0.5 \textwidth]{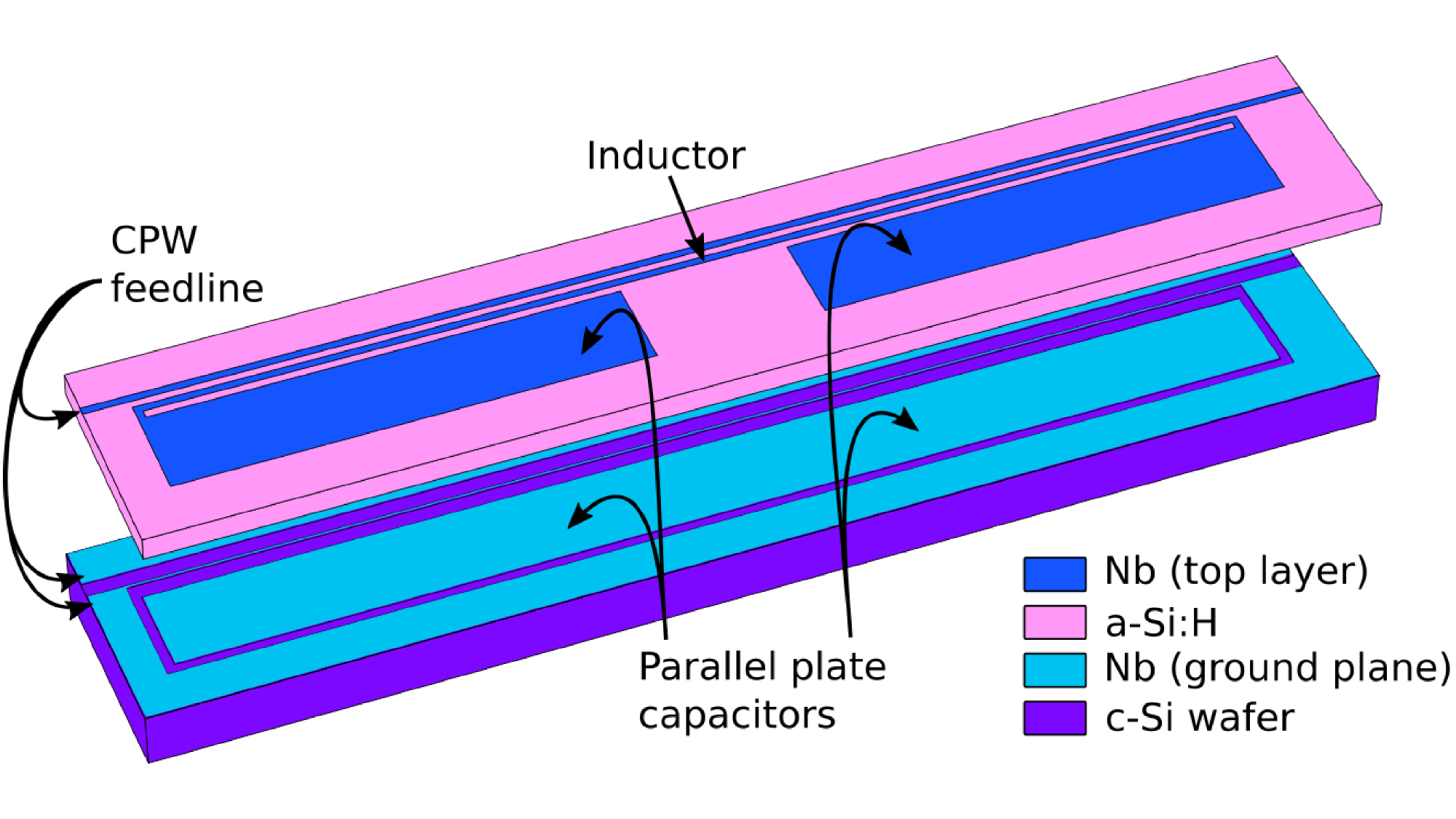}

\includegraphics[width=0.5 \textwidth]{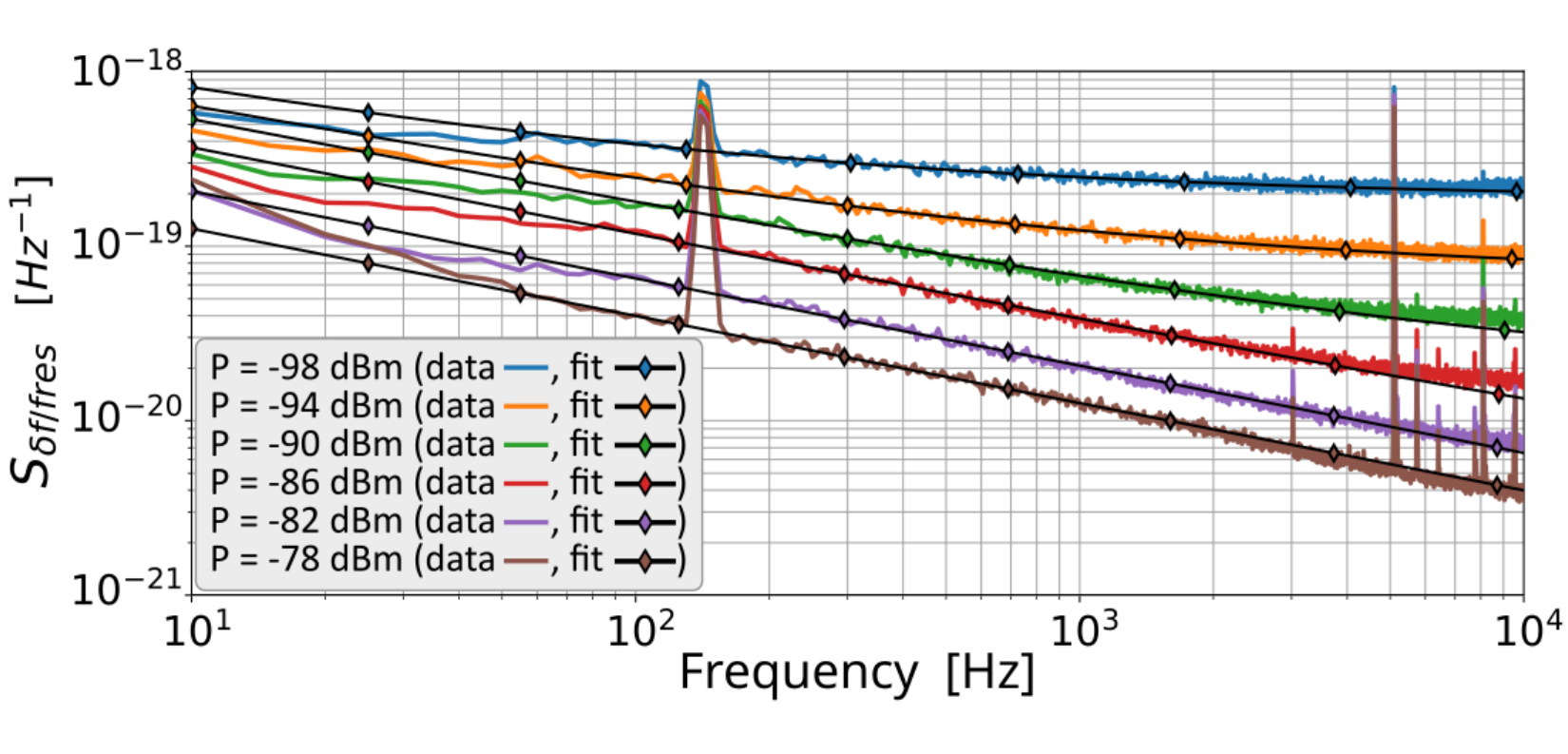}
\caption{\protect\brep{Top: Design of the }\protect\mrep{a-Si:H PPC TLS noise }characterization devices\protect\bsout{ made by Defrance\mbox{\cite{defrance-c}}}\protect\mrep{\ used in\mbox{\cite{defrance-c,defrance}}}. Bottom: \protect\bsout{TLS noise performance of a reference Nb superconducting resonator incorporating 
a-Si:H dielectric PPC}\protect\mrep{Example TLS noise measurements from}~\cite{defrance}.  Noise \protect\mrep{power spectral densities (}PSDs\protect\mrep{)} are shown for \protect\gsout{an 847 MHz }
\protect\mrep{a }resonator\protect\mrep{\ with $f_{res} = 847$~MHz}. 
\protect\mrep{The d}\protect\gsout{D}ata were fit \protect\gsout{with }\protect\mrep{to }$a\protect\gsoutmath{f^{-0.5}}\protect\mrep{/\sqrt{f}}+b$ to capture both TLS and white noise. \protect\mrep{The p}\protect\bsout{P}\protect\brep{eak at 120 Hz is \protect\bsout{a result of }power line pickup. }
Bottom was reprinted from Defrance, et al., Appl. Phys. Lett., vol. 126, p. 122904, Mar. 2025,
with the permission of AIP Publishing \cite{defrance}. Top was adapted with permission from Defrance et al., Phys. Rev. Mater. 8, 035602 (2024). Copyright 2024 APS \cite{defrance-c}.
}
\label{fig:tls_noise_performance}
\end{figure}

\rsout{Our prior measurements of the TLS noise of our a-Si:H PPCs in the 0.1-10~kHz range\mbox{\cite{defrance}}, reproduced here in Fig.~\ref{fig:tls_noise_performance}} \bsout{along with the detector design for the TLS measurement}\rsout{, enable a prediction for TLS noise in our MS-PPC-LEKIDs at lower frequencies and in the operating conditions used.} 
\mrep{In~\cite{defrance}, we characterized the TLS noise of our a-Si:H PPCs in the 0.1-10~kHz range.  We reproduce here in Fig.~\ref{fig:tls_noise_performance} the design of the devices used for those measurements as well as the measurements themselves.  We scale from these measurements to predict the TLS noise amplitude in our Al/a-Si:H MS-PPC-LEKIDs.} 

The scaling relation \mrep{we assume} is
\ifthenelse{\value{showedits}=0}{}{\begin{equation}
   \gsoutmath{S_{TLS} = S^{r}_{TLS}\left(\frac{T}{246 \text{ mK}}\right)^{-0.8}\frac{A}{A_r}\frac{E}{E_r},} \nonumber 
\end{equation}}\begin{equation}
   \mrep{S_{TLS}(\text{1 kHz}) = S^{r}_{TLS}(\text{1 kHz})\,\frac{A_r}{A}\frac{E_r}{E}\left(\frac{246 \text{ mK}}{T}\right)^{0.8}}
    \label{eq:scaling}
\end{equation}
where $A$ is the capacitor area, $E$ \mrep{is }the magnitude of the electric field $\vec{E}$, and the subscript $_r$ denotes reference values from our prior work~\cite{defrance}.  \mrep{The electric field and temperature scaling laws were established in~\cite{gaonoise1} and \cite{kumar_noise}.  The expectation that the noise of individual TLS add incoherently (in quadrature) implies linear volume scaling, which we convert here to area scaling because we use the same film thickness as in \cite{defrance}. Table~\ref{tbl:val_table} shows the values we use to evaluate Eq.~\ref{eq:scaling}:
the reference device parameters and TLS noise value, the area of the device considered here, the $E$ field values at which our measurements were made, and the TLS noise expectation for our measurements.}\gsout{\ That work used 
broken power law fits to the TLS data,}
\ifthenelse{\value{showedits}=0}{}{\begin{equation}
\gsoutmath{S(f) = 
    \begin{cases}
        af^{-1} + b, & f \leq 100 \ \text{Hz}, \\
        cf^{-0.5} + d, & f > 100 \ \text{Hz},
    \end{cases}
}
    \label{eq:tls_power_law}
\end{equation}
}\gsout{consistent with prior TLS studies that find a transition from $f^{-0.5}$ to $f^{-1}$ scaling near 100 Hz\mbox{\cite{defrance}}.}

\bsout{For EQ. \ref{eq:scaling}, the reference and measurement values for the area and electric fields are shown in FIG. \ref{fig:val_table}.
The fitted values from the reference measurements, as well as the values fitted onto the dark noise data is shown in FIG. \ref{fig:dark_noise_fits}. Both the optical and dark noise shown in FIGs. \ref{fig:dark_load} and \ref{fig:optical_load}, was taken at 245 mK, with a readout power at the device of $\sim$ 80 dBm.}

\protect\ifthenelse{\value{forreferee}=0}{\ifthenelse{\value{showedits}=0}{}{
\begin{figure}[t]
\centering
\footnotesize
\bsout{
\begin{tabular}{lcc}
\hline\hline
KID & Area ($\mathrm{m^2}$) & Electric Field ($\mathrm{V/m}$) \\ 
\hline
Reference TLS KID
  & $0.30\times10^{-6}$
  & $130$ \\
NEW-MUSIC KID
  & $0.55\times10^{-6}$
  & $1519$ \\
\hline\hline
\end{tabular}
}
\caption{\protect\bsout{Comparison of KID capacitor area and electric field strength for values used in EQ. \ref{eq:scaling}.}}
\label{fig:kid_area_field}
\label{fig:val_table}
\end{figure}
}{}}

\begin{table}[t]
\centering
\footnotesize
\mrep{
\begin{tabular}{lccc}
\hline\hline
Device & A (m$^2$) & E (V/m) & $S_{TLS}$(1~kHz) \\
\hline
Reference\mbox{\cite{defrance}}
  & $0.30\times10^{-6}$
  & $130$ 
  & $2.5\times 10^{-19}$ \\
\multicolumn{4}{l}{Al/a-Si:H MS-PPC-LEKID (this work and\mbox{\cite{SPIE}}} \\ 
\hspace{12pt}Dark
  & $0.55\times10^{-6}$
  & $1713$ 
  & $8.6 \times 10^{-21}$ \\
\hspace{12pt}Optical
  & $0.55\times10^{-6}$
  & $990$ 
  & $14.8 \times 10^{-20}$ \\
\hline\hline
\end{tabular}
}
\protect\mrep{\vspace{6pt}\caption{Reference values for Eq.~\ref{eq:scaling} from \cite{defrance} and values for the device studied here.}}
\label{tbl:val_table}
\end{table}

\mrep{It will also be necessary to scale from the 1~kHz reference measurement frequency to the measurement frequencies of interest here.  Over the 0.1--10~kHz range studied in\mbox{\cite{defrance}}, a $1/\sqrt{f}$ dependence was observed, so that is one extrapolation model we use.  A survey of the literature in\mbox{\cite{defrance}} suggests the power law may change to $1/f$ below 100~Hz.  We thus also consider an extrapolation using a broken power law: $1/f$ below 100~Hz and $1/\sqrt{f}$ above 100~Hz.}

\ifthenelse{\value{showedits}=0}{}{\red{\section{\sout{Noise Measurements and Results}}}\addtocounter{section}{-1}\vspace{-12pt}}
\mrep{\section{Noise Measurements under Dark Conditions\label{sec:dark_noise}}}

\begin{figure}[t]
\centering
\includegraphics*[width=0.5 \textwidth,viewport=0 0 600 592]{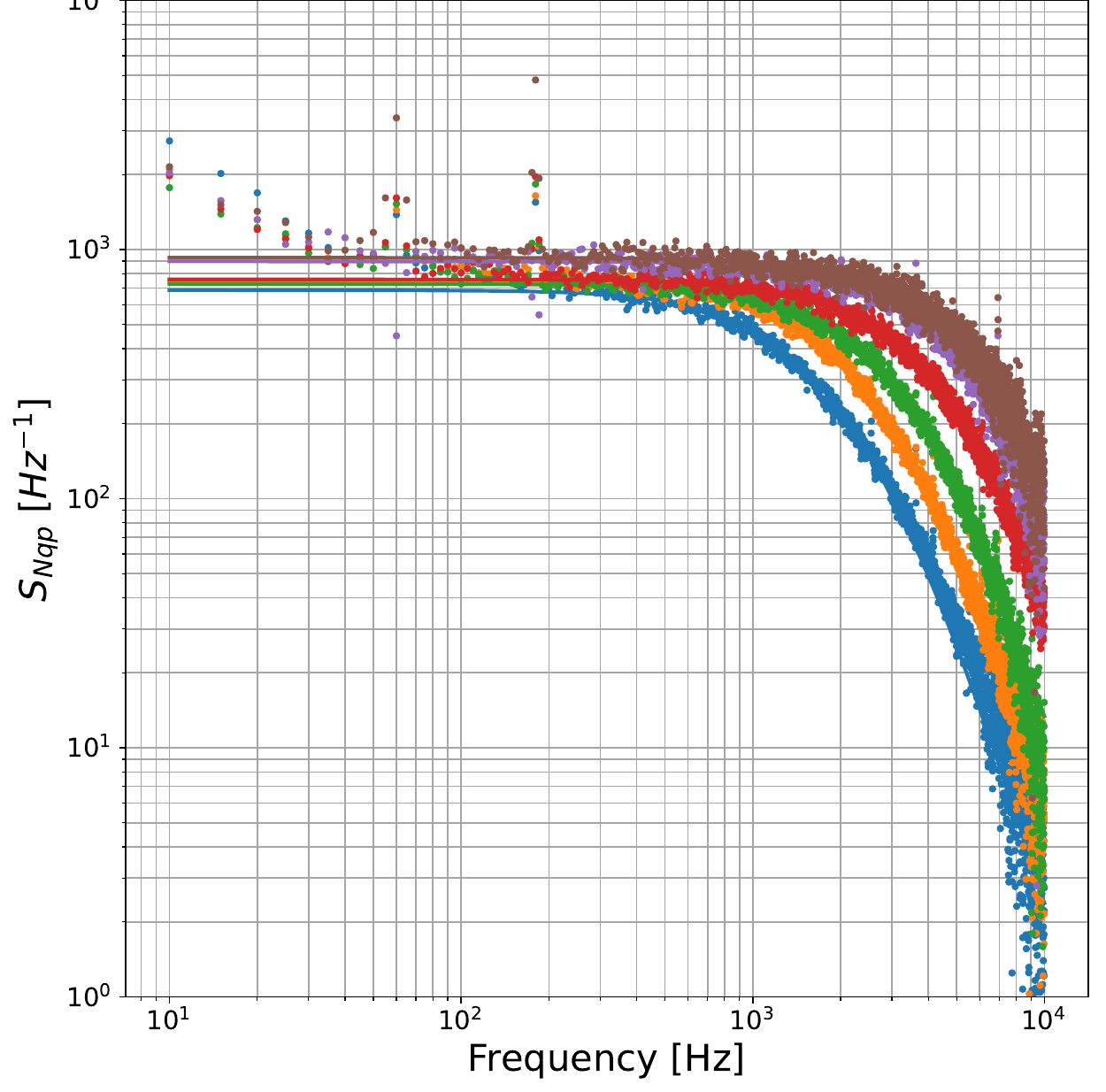}
\protect\mrep{\caption{Dark noise data for Al/a-Si:H MS-PPC-LEKID with $f_{res} = 255.83$~MHz in $S_{N_{qp}}$ units, (quasiparticles)$^2$/Hz, for a range of bath temperatures. Prior to conversion to $S_{N_{qp}}$, off-resonance noise PSDs were subtracted from the on-resonance PSDs to remove the sub-dominant electronics white noise contribution.  The temperature varies from 250 to 350~mK in steps of 20~mK in the sequence blue-orange-green-red-purple-brown.  The text explains that generation-recombination noise is dominant from $\approx$ 100~Hz to the high-frequency rolloff in these data.}}
\label{fig:dark_load_Tvary}
\end{figure}

\begin{figure}[t]
\centering
\includegraphics[width=0.5 \textwidth]{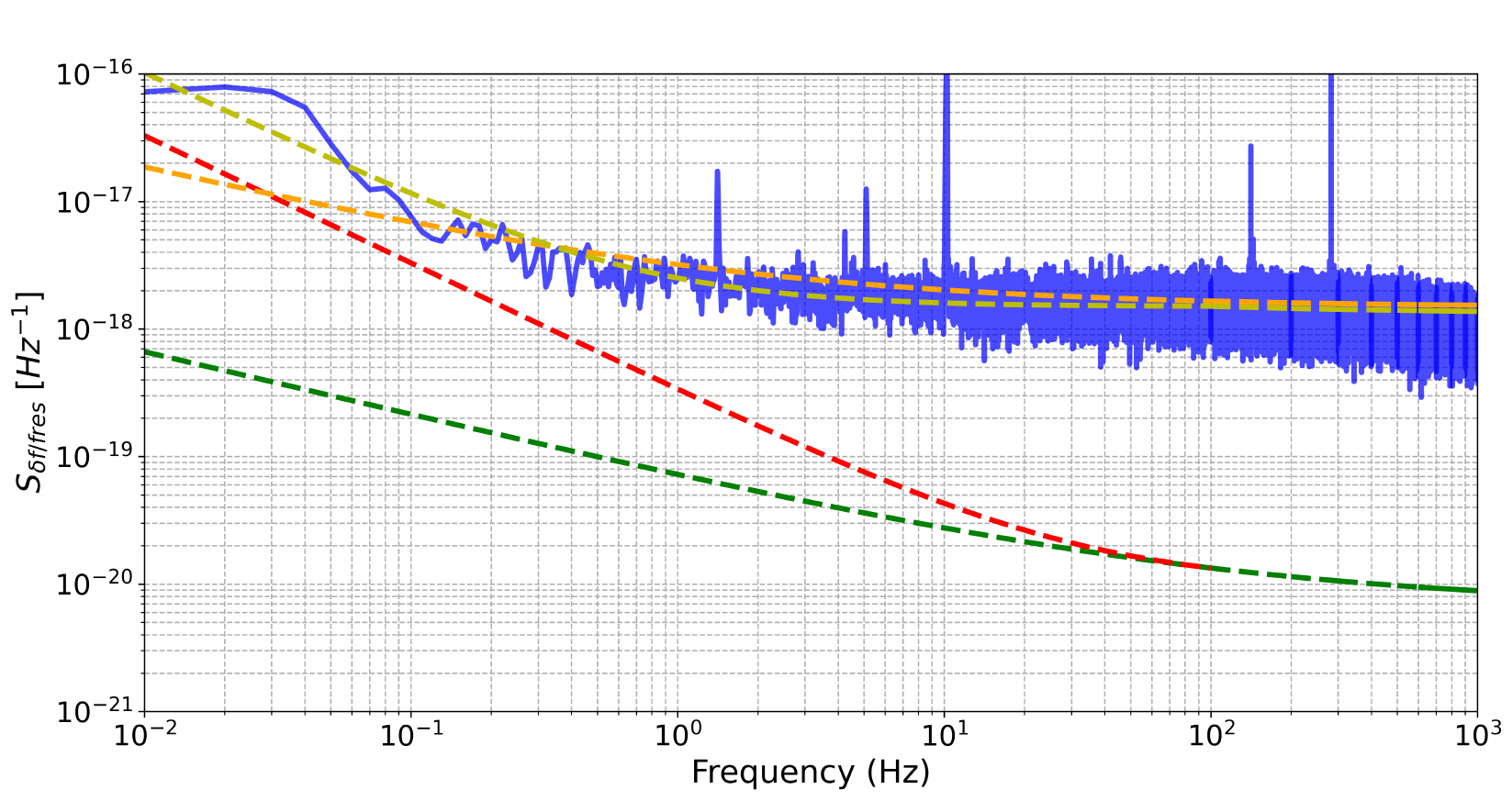}
\caption{
Dark noise data \protect\gsout{and }\protect\mrep{for Al/a-Si:H MS-PPC-LEKID with $f_{res} = 255.83$~MHz in $S_{\delta f/f_{res}}$ units, (Hz/Hz)$^2$/Hz = 1/Hz, with }TLS noise models.  Blue: Noise PSD under dark conditions\protect\gsout{ for an Al MS-PPC-LEKID ($f_r = 253.83$ MHz)}\protect\mrep{, now measured with the two-tone IQ mixer setup (\S\ref{sec:two-tone} and Fig.~\ref{fig:Two Tone Setup}), which is more stable at low frequencies than the single-tone IQ mixer setup used for the data in Fig.~\ref{fig:dark_load_Tvary}}. 
\protect\bsout{Red and green}\protect\mrep{Green and red}: 
extrapolated scaling of TLS noise measured for a-Si:H PPCs in \cite{defrance} to this resonator's 
PPC area and electric field for the feedline readout power used \protect\mrep{(see \S\ref{sec:tls} for details of the scaling and \S\ref{sec:dark_noise} for how the electric field is calculated)}. 
Since \cite{defrance} only measured noise above 100 Hz, below 100~Hz 
we show both 
the\protect\bsout{ Eq.~2 broken power law and a simple $f^{-1}$ extrapolation 
(
red and green, respectively).}
\protect\mrep{ $1/\sqrt{f}$ single power law (green) and the $1/\sqrt{f} \to 1/f$ broken power law (red) models described in \S\ref{sec:tls}.}
\protect\gsout{Mustard}\protect\brep{Orange} and yellow-green: approximate fits of these two models \protect\mrep{plus white noise, Eq.~\ref{eq:tls_power_law},} 
to the measured dark noise PSDs, 
from which we derive approximate upper limits on TLS noise\protect\mrep{ (see \S\ref{sec:dark_noise})}.  It is not certain the low-frequency rise is TLS noise, and there is good reason to think it is not \protect\gsout{(see text)}\protect\mrep{and that it may be reduced in future measurements$^\text{\protect\ref{ftn}}$}.
}
\label{fig:dark_load}
\end{figure}

Noise was measured under \gsout{both }dark \gsout{and optically loaded }conditions\gsout{.}\mrep{\ for an Al/A-Si:H MS-PPC-LEKID with $f_{res} = 255.83$~MHz.  Two datasets were taken at different times and using different readout systems with differing low-frequency stability.}\bsout{Noise was measured on the same device, denoted as Be231102p2 and the same resonance ($f_0 = 255.83$MHz), for both measurements. In both Fig.~\ref{fig:tls_noise_performance} and Fig.~\ref{fig:optical_load}, the noise data were acquired using an IQ measurement setup. The two measurement setups were nearly identical; however, for the dark data shown in Fig.~\ref{fig:tls_noise_performance}, the two tone system was used, while for the optical data shown in FIG. ~\ref{fig:optical_load}, the one tone version of the system was used. Loop back testing was done on both systems, to guarantee that system noise significantly below frequency noise.}\gsout{The optical loaded data was taken with 
$T_{\text{Load}} = 77 \text{K} +  T_{\text{exc}}$}\ifthenelse{\value{forreferee}=0}{\ifthenelse{\value{showedits}=0}{}{\blue{\cancel{\footnotemark}}\footnotetext{\bsout{$T_{\text{exc}} $ refers to the excess load due to emission from dewar. It is fit based on measurements of $\delta f_r/f_r$ under blackbody load at different temperatures. Temperatures are Rayleigh-Jeans temperatures for the blackbody load.  Optical loading was done using a blackbody filled with LN2, placed such that it filled view of the cryostat window.}}}}{}
\gsout{$\approx 180$K. }Representative 
noise power spectral densities (PSDs) \gsout{in fractional frequency units, $S_{\delta f/f}$ }are shown in Figs.~\mrep{\ref{fig:dark_load_Tvary} and }\ref{fig:dark_load}\gsout{ and \ref{fig:optical_load}}.  


\mrep{The first data set, reported on initially in~\cite{SPIE}, included measurements of fractional frequency noise, $S_{\delta f/f_{res}}$, at a range of temperatures.  For reasons that will become clear below, they were converted to and presented in $S_{N_{qp}}$ units, (quasiparticles)$^2$/Hz, using
\begin{equation}
S_{N_{qp}} = \frac{V^2}{(\alpha\, \kappa_2)^2}\,S_{\delta f/f_{res}}   \label{eq:Sdff_to_SNqp}
\end{equation}
where $V \approx 3500$~$(\mu\text{m})^3$ is the inductor volume, $\alpha$ is the kinetic inductance fraction, and $\kappa_2(T) = \left. \partial \left(\sigma_2/|\sigma(T=0)|\right)/\partial n_{qp} \right|_T$ is the standard factor converting changes in quasiparticle density, $n_{qp} = N_{qp}/V$, to changes in the imaginary part of the complex conductivity (normalized to the magnitude of the complex conductivity at zero temperature) and thus fractional frequency shift.  $\kappa_2(T)$'s dependence on temperature and quasiparticle density are very weak\mbox{\cite{Zmuidzinas}}.  The data were taken using a standard single-tone IQ mixer setup whose stability degraded below tens of Hz, as is evident from the data.  We thus neglect the rise below 100~Hz and focus on the flat noise level above 100~Hz.}

\mrep{Fig.~\ref{fig:dark_load_Tvary} shows that $S_{N_{qp}}$ from those data is approximately independent of temperature while the 3~dB roll-off frequency moves up with temperature.  These behaviors are consistent with expectations for generation-recombination (GR) noise: the flat level should have temperature-independent value $S_{N_{qp}}^{GR} = 4\,\tau_{qp}\,N_{qp} \approx 2\,V/R$ ($R \sim 10$~$(\mu\text{m})^3$/s is the recombination constant) while the 3~dB roll-off frequency should by given by the reciprocal of $ 2\,\pi$ times the quasiparticle lifetime, $(2\,\pi\tau_{qp})^{-1} \approx 2 R\,N_{qp}/V/2\,\pi$, which increases with temperature as $N_{qp}$ increases with temperature.  In~\cite{SPIE}, we thus concluded from these data that our Al/a-Si:H MS-PPC-LEKID design's noise PSD is dominated by GR noise above 100~Hz.}

\mrep{The second data set, shown in Fig.~\ref{fig:dark_load}, was taken at a later date after the development of the two-tone setup described in \S\ref{sec:two-tone} and Fig.~\ref{fig:Two Tone Setup}.  While, as noted above, no ``cleaning'' was performed, it is clear from the data that the two-tone setup had better low-frequency stability than the single-tone setup.  The second data set's goal was thus to use the more stable readout electronics to probe the low-frequency noise of the Al/a-Si:H MS-PPC-LEKID design, in particular to determine the frequency down to which GR noise dominates over TLS noise.}

\rsout{Under dark conditions, the noise PSDs above $\approx 0.1$ ~Hz }\gsout{are dominated above $\approx$0.1~Hz by} \bsout{are consistent with }\rsout{gen\-er-a\-tion\-recombination noise.  (This point has been verified by checking its temperature dependence\mbox{\cite{SPIE}}.)}

\mrep{In Fig.~\ref{fig:dark_load}}, \gsout{W}{w}e \mrep{therefore} overlay the \bsout{expected }TLS noise \mrep{expected }based on \bsout{scaling using }the Eq.~\ref{eq:scaling} scaling for \bsout{the known KID area and for the known feedline readout power and the calculated stored energy stored in the KID (see, e.g.,\mbox{\cite{defrance-c}} for the conversion)}\mrep{the capacitor area and electric field and both the $1/\sqrt{f}$ single power law and the $1/\sqrt{f} \to 1/f$ broken power law described in \S\ref{sec:tls}.  These data were taken at $T = 239$~mK with feedline power $P_{feed} \approx$~--80~dBm and resonator coupling and internal quality factors \brep{$Q_c = 20 $k and $Q_i = 240$k}, and we calculate the electric field from the stored energy $W$ and feedline power $P_{feed}$ using
\begin{equation}
    \label{eq:field}
    E = \sqrt{\frac{2\,W}{\varepsilon_{_r}\,\varepsilon_{_0}\,A\,d}} 
      = \sqrt{\frac{2}{\varepsilon_{_r}\,\varepsilon_{_0}\,A}} \sqrt{\frac{P_{feed}}{\pi f_{res}} \frac{Q_r^2}{Q_c}}
\end{equation}
where $\varepsilon_{_0}$ is the permittivity of free space, $\varepsilon_{_r} \approx 12$ and $d = 800$~nm are the relative permittivity and thickness of the a-Si:H dielectric, and $Q_r = (Q_c^{-1} + Q_i^{-1})^{-1}$ is the resonator quality factor.\footnote{\mrep{Note that the precise value of $\varepsilon_{r}$ is unimportant for the scaling as long as it is common to the measurements in~\cite{defrance} and here: it is present in both $E_r$ and $E$ in Eq.~\ref{eq:scaling}.}}}  \bsout{We plot the scaled TLS noise assuming the given broken power law and assuming extrapolation of the $f^{-0.5}$ power law at all frequencies.}

The scaling estimates 
predict TLS noise levels below the noise PSD observed. 
Thus, while we cannot explicitly check the extrapolation of the TLS noise measurements from~\cite{defrance}, we can\bsout{ set upper limits by fitting the $f \leq 100$~Hz form to the observed data,} 
conservatively \bsout{assuming }\mrep{assume }the rise in noise at low frequency is entirely due to TLS\bsout{.}\footnote{We have no definite evidence it is TLS\mrep{\ noise}.  In fact, the appearance of a similar rise in the dissipation-direction noise PSD (not shown here), which should not be subject to TLS noise, suggests the rise is correlated electronics noise.  \gsout{Future measurements will benefit from removal of correlated noise cleaning}\mrep{It may be reduced when the correlated-noise-removal capability of the two-tone IQ mixers setup is used}.\label{ftn}}\mrep{\ and fit the following two forms (for the two TLS power laws combined with the observed white noise) to the Fig.~\ref{fig:dark_load} data:
\begin{equation}
    S(f) = 
    \begin{cases}
        a \left[f/(\text{1~Hz})\right]^{-1/2} + b \\
        c \left[f/(\text{1~Hz})\right]^{-1} + d\\
    \end{cases}
    \label{eq:tls_power_law}
\end{equation}
Table~\ref{tbl:dark_noise_fits} shows the fitted parameters. }We thus obtain \bsout{an }approximate upper limit\mrep{s at $f = 0.1$~Hz} of \bsout{3.5 times the broken power-law model and }56 times the \bsout{extrapolated $f^{-0.5}$}\mrep{$1/\sqrt{f}$} model \mrep{and 3.5 times the broken power-law model\bsout{ at $f= 0.1 \text{Hz}$}.}

\ifthenelse{\value{showedits}=0}{}{\begin{figure}[t]
\centering
\footnotesize
\bsout{
\begin{tabular}{lcc}
\hline\hline
 & Slope (a \& c) & Intercept (b \& d) \\
\hline
$f^{-1}$ Dark Noise
  & $1.01\times10^{-18}$
  & $1.75\times10^{-18}$ \\
$f^{-0.5}$ Dark Noise
  & $1.962\times10^{-18}$
  & $1.316\times10^{-18}$ \\
$f^{-1}$ Scaled TLS Noise
  & $9.56\times10^{-18}$
  & $9.67\times10^{-18}$ \\
$f^{-0.5}$ Scaled TLS Noise
  & $1.9125\times10^{-18}$
  & $1.978\times10^{-18}$ \\
\hline\hline
\end{tabular}
}
\caption{\protect\bsout{Slope and intercept values for the best fits to the dark noise data  and the reference TLS data. Slope and Intercept refer to a,b and c,d in EQ. \ref{eq:tls_power_law} for $f^{-1}$ and $f^{-0.5}$.  We also show the normalization of the models scaled from\protect\mbox{\cite{defrance}}.}}
\label{fig:dark_noise_fits}
\end{figure}
}

\begin{table}[t]
\centering
\footnotesize
\mrep{
\begin{tabular}{lcc}
\hline\hline
 &  a \& c & b \& d \\
\hline
$1/\sqrt{f}$ Scaled TLS Noise
  & $6.5\times10^{-20}$
  & 0 \\
$1/\sqrt{f} \to 1/f$ Scaled TLS Noise
  & $3.2\times10^{-19}$
  & 0 \\
$1/\sqrt{f}$ Fitted Noise
  & $2.0\times10^{-18}$
  & $1.3\times10^{-18}$ \\
$1/f$ Fitted Noise
  & $1.0\times10^{-18}$
  & $1.8\times10^{-18}$ \\
\hline\hline
\end{tabular}
}
\protect\mrep{\vspace{6pt}\caption{Parameters of Eq.~\protect\ref{eq:tls_power_law} TLS models shown in Fig.~\protect\ref{fig:dark_load}.  \\ Top two rows: normalization of the models scaled from\protect\mbox{\cite{defrance}} (no white noise term).  Bottom two rows: Best fits to the dark noise data (now including white noise).}}
\label{tbl:dark_noise_fits}
\end{table}

\mrep{\section{Extrapolation to Noise Under Optical Load}\label{sec:opt_load}}

\gsout{Under optical load, the noise has only been reliably measured above 100 Hz (those measurements, previously reported in\mbox{\cite{SPIE}}, did not use the two-tone setup.}  
\mrep{Noise measurements under optical load for the same resonator as studied in \S\ref{sec:dark_noise} were also reported in\mbox{\cite{SPIE}}, taken at $T = 241$~mK\footnote{\mrep{The small difference in resonator temperature between the dark and optical data has negligible impact given the weak $T$-dependence of photon and GR noise in $S_{N_{qp}}$ units (via $n_{qp,th}^2/n_{qp}^2$ in Eq.~\ref{eq:psd_gr_photon_Nqp}) and of the conversion between $S_{N_{qp}}$ and $S_{\delta f/f_{res}}$ (Eq.~\ref{eq:Sdff_to_SNqp}).}} and $P_{feed} \approx$~--80~dBm like our dark measurements, but with 
$T_\text{Load} = 77\ \text{K} + T_\text{exc} \approx 180$~K.\footnote{\mrep{$T_\text{Load}$ values are equivalent Rayleigh-Jeans temperatures for a blackbody load in front of the cryostat window.  These measurements were done with a blackbody held at LN$_2$ temperature.  $T_{\text{exc}}$ refers to optical loading in excess of that presented at the cryostat window.  It is believed to primarily be emission from the UHWMPE optical window and PTFE thermal filters at the 4~K and 50~K states of the pulse-tube-cooled cryostat.}} Those measurements used the same single-tone IQ mixer setup as the dark noise measurements in Fig.~\ref{fig:dark_load_Tvary} and thus were limited by low-frequency electronics noise below 100~Hz.}
\mrep{Above 100~Hz, t}\gsout{T}hey \mrep{were flat (up to the quasiparticle lifetime roll-off) and }showed a dependence on optical load consistent with domination by \gsout{GR+photon }\mrep{photon and GR }noise\mrep{: in $S_{N_{qp}}$ units, the expectation is (Eq.~2 of \cite{SPIE}, slightly modified)
\begin{multline}
S^{\gamma+GR}_{N_{qp}} = \\ 
\frac{V}{R} \left[
\eta_{pb} \left( \frac{h\,\nu}{2\,\Delta} + \frac{\eta_{op}\,k_{_B}\,T_\text{Load}}{2\,\Delta}\right) 
+ 1 + \frac{n_{qp,th}^2}{n_{qp}^2} \right]
\label{eq:psd_gr_photon_Nqp}
\end{multline}
where $\nu$ is the observing frequency (220~GHz for this data set\footnote{\mrep{Though not relevant here, we note the spectral bandwidth was $\Delta \nu \approx 45$~GHz, set by an on-chip bandpass filter. See\mbox{\cite{SPIE}} for a description of the bandpass filter and measurements.}}), $2\,\Delta$ is the superconductor's pair-breaking energy, $h$ and $k_{_B}$ are Planck's and Boltzmann's constants, $\eta_{op}$ is the optical efficiency between the optical load and the detector, $\eta_{pb}$ is the efficiency with which photon energy breaks pairs (unity at $h\,\nu = 2\,\Delta$ but decreasing to $\approx 0.42$\mbox{\cite{guruswamy}} for larger $\nu$ due to losses to phonon emission and escape), and $n_{qp,th}$ and $n_{qp}$ are the thermally generated and total (optically and thermally generated) quasiparticle densities at $T$.  Indeed, those data (Fig.~14 of}\mbox{\cite{SPIE}}\gsout{.}\mrep{) showed a linear dependence on $T_\text{Load}$ consistent with the Bose photon noise (second term), and the offset at $T_\text{Load} = 0$ was larger than the GR noise observed in the dark measurements (last two terms), as expected when the $T_\text{Load}$-independent shot noise term (first term) is included. Photon and GR noise are inherently white, so we }\gsout{We thus }extrapolate \gsout{this measured noise down to 0.01 Hz in the figure}\mrep{the $>$100~Hz white noise level for $T_\text{Load} \approx 180$~K (the lowest $T_\text{Load}$ applied) to low frequencies for comparison to a scaling of the dark TLS noise measured here}.   

\begin{figure}[t]
\centering
\includegraphics[width=0.5 \textwidth]{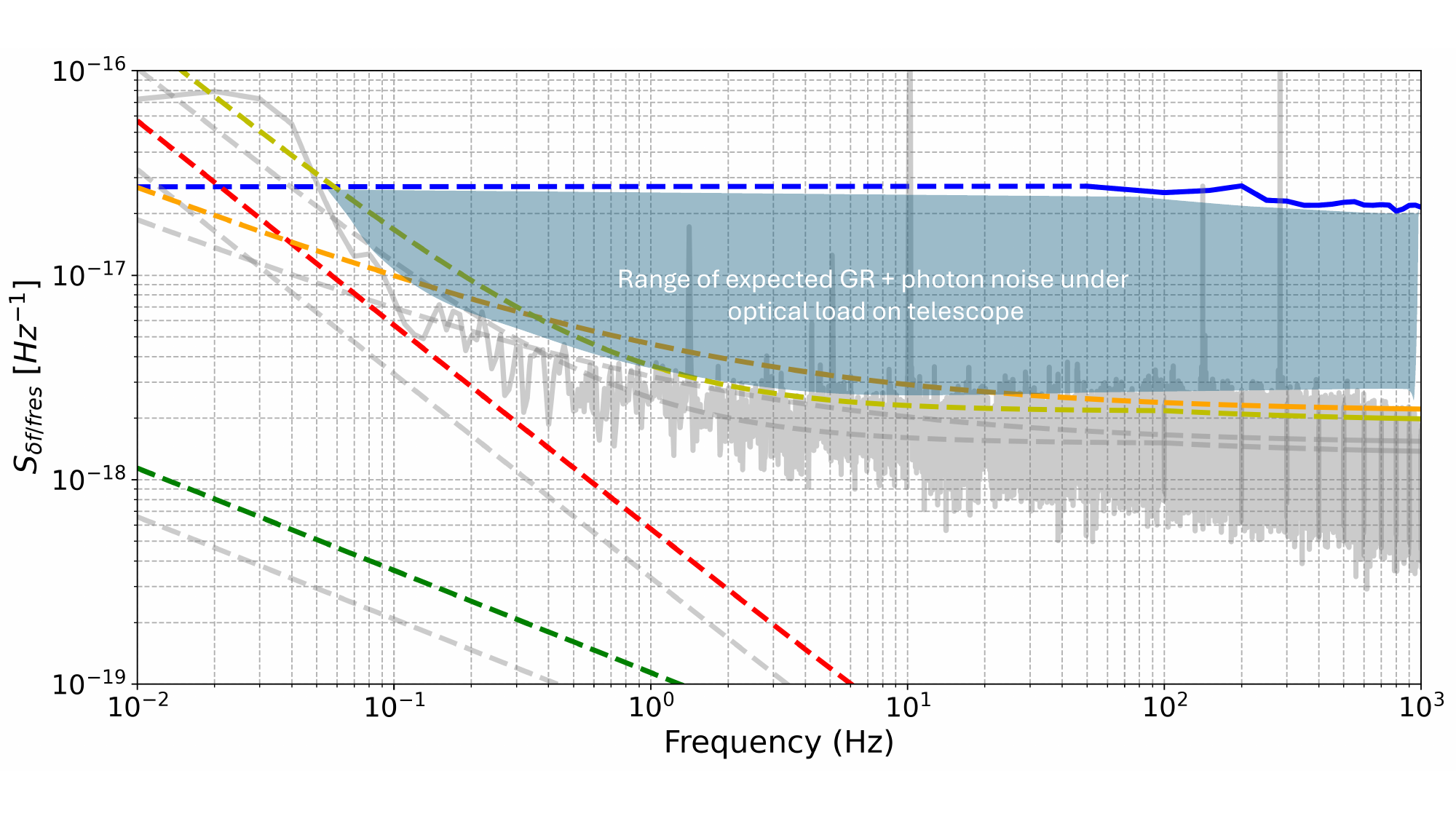}
\caption{\protect\gsout{Noise PSD }\protect\mrep{Noise data }under optical load\protect\gsout{,} compared to dark noise PSD and TLS noise estimates. 
Blue solid: \protect\gsout{GR+photon }noise measured above 100~Hz under $T_\text{Load} = 180$~K. These data were taken with coarser frequency resolution and more averaging than the dark data in Fig.~\protect\ref{fig:dark_load} included here in grey and so do not show similar scatter. \protect\mrep{The text provides evidence that this noise is dominated by photon and GR noise.}
Blue dashed: \protect\mrep{flat }extrapolation of that noise level to low frequency\protect\bsout{ assuming noise is flat}\protect\mrep{, appropriate for photon and GR noise}.  Grey solid: dark noise PSD\protect\mrep{\ from Fig.~\ref{fig:dark_load}}.  Grey dashed: TLS noise model curves from Fig.~\ref{fig:dark_load}, unmodified.  Colored dashed: the same TLS noise model curves \protect\gsout{extrapolated to the operating conditions under optical load (i.e., scaling by electric field magnitude)}\protect\mrep{scaled to the PPC electric field magnitude under optical load (see \S\ref{sec:opt_load} for details)}: they increase because the \protect\gsout{stored power }\protect\mrep{electric field }decreases.\protect\gsout{ Discussion is provided in the text.}\protect\mrep{\S\ref{sec:opt_load} explains how the measured dark noise (dominated by GR noise) and the extrapolated measured noise under optical load (dominated by photon and GR noise) bracket the noise expected under telescope on-sky optical load and how the unscaled (gray dashed) and scaled (colored dashed) TLS models bracket the expected TLS noise under those conditions.}
}
\label{fig:optical_load}
\end{figure}

We overlay the noise PSD from the dark data\mrep{\ measured in this work}.  Because \gsout{the optical load was $\approx$180~K, and }we expect the optical load on\gsout{ the}\mrep{-}sky \mrep{at the telescope }will be 40--150~K (depending on spectral band)~\cite{SPIE},\mrep{\ the dark and optical data bracket the expected operating conditions.}\gsout{ and because the KID responsivity $d(\delta f/f)/dP_{opt}$ decreases with increasing load, the photon noise under on-sky conditions will likely be between the dark GR noise PSD and the optical GR+photon noise PSD.  A collusion of the dependence of \rsout{responsibility }\bsout{\brep{responsivity }}and photon noise on loading could possibly cause the photon noise in $S_{\delta f/f}$ to exceed the value measured at $T_\text{Load} = 180$~K; if so, our analysis below is conservative.}  \mrep{Recall that the PSD of photon and GR noise in quasiparticle number and fractional frequency are related by Eq.~\ref{eq:Sdff_to_SNqp}, with the proportionality factor consisting of terms with weak or no temperature dependence.  $S^{\gamma+GR}_{\delta f/f_{res}}$ is thus also an almost monotonic function of optical load with the same terms as $S^{\gamma+GR}_{N_{qp}}$ (Eq.~\ref{eq:psd_gr_photon_Nqp}).  The second (Bose) term is linear in $T_\text{Load}$, while the first (shot) term generally more than compensates the reduction in the fourth term (part of GR) when an optical load is introduced ($n_{qp} = n_{qp,th}$ when $T_\text{Load} = 0$~K while $n_{qp} > n_{qp,th}$ when $T_\text{Load} > 0$~K).  Certainly, over the roughly one order of magnitude increase in $S_{\delta f/f_{res}}$ in Fig.~\ref{fig:optical_load} between dark and $T_\text{Load} \approx 180$~K conditions, the Bose term dominates, and thus, on sky at the telescope, $S_{\delta f/f_{res}}$ will be in the shaded region in the figure.}

Next, \gsout{we scale }\mrep{in Fig.~\ref{fig:optical_load}, the colored dashed lines show the scaling of }both the TLS noise measurements from~\cite{defrance} and the upper limit from the dark data \mrep{(\S\ref{sec:dark_noise}) }to the optically loaded operating conditions \mrep{(\textit{i.e.}, reduced $E$ due to lower $Q_i$ under load; recall that $P_{feed} \approx$~--80~dBm is unchanged) for comparison }\gsout{ and compare }to the extrapolated \gsout{GR+photon }\mrep{photon and GR }noise. \gsout{These scalings are the various other colored curves in Fig.~\ref{fig:optical_load}. }\mrep{We also show the curves from Fig.~\ref{fig:dark_load} without any scaling in grey. }\gsout{We see }\mrep{Fig.~\ref{fig:optical_load} implies }that\mrep{,} even \mrep{for }the most conservative case \mrep{(\textit{i.e.}, highest TLS noise,} \gsout{--- }the broken power law upper limit\gsout{ ---}\mrep{)} \gsout{implies }photon \mrep{and GR }noise dominate\gsout{s} down to 0.1~Hz under 180~K optical load. \mrep{We argued above that, f}\gsout{F}or\gsout{intervening} optical loads\gsout{,} between dark and 180~K, the \mrep{sum of the }photon \mrep{and GR }noise\mrep{s} will \gsout{likely }lie \mrep{in the shaded region }between the \gsout{180~K noise and the dark noise}\mrep{dark and 180~K noise.}\gsout{, while}\mrep{\ Under the same conditions, }the TLS noise will lie between the colored and companion grey curves shown \mrep{in }Fig.~\ref{fig:optical_load}\mrep{\ because $Q_i$ and thus $E$ will lie between the dark and 180~K values at $P_{feed} \approx$~--80~dBm}.  The crossing point of the \gsout{photon noise and }scaled TLS noise \textit{upper limits} \mrep{with the photon and GR noise }may thus, conservatively, \gsout{reach into the }\mrep{be as low as }tenths of a Hz.  The crossing point may in fact be lower if the low-frequency rise seen in the dark data is removable electronics, rather than TLS, noise.$^\text{\ref{ftn}}$  If the prediction \gsout{from }from~\cite{defrance} is \mrep{instead }correct, the crossing point is 0.1-0.2~Hz (broken power law) or \gsout{lower frequency ($f^{-1}$ }\mrep{below ($1/\sqrt{f}$ }power law).


\section{Conclusion}



In this paper, we measured the low-frequency noise of \mrep{the }Al/a-Si:H MS-PPC-LEKID\gsout{s of the NEW-MUSIC} design \mrep{presented in\mbox{\cite{SPIE}} }under dark conditions\rsout{,observing GR noise down to 0.1 Hz}\bsout{\ observing a flat frequency noise excess indicative of GR noise}\mrep{\ and, in combination with prior data, demonstrated that GR noise is dominant over TLS noise}\brep{\ down to about 0.1 Hz}.  The upper limits we can set on TLS noise are a factor of 
3.5 to 56 above the expectation based on scaling of TLS noise measurements of a-Si:H PPCs in~\cite{defrance}\gsout{ but consistent with those predictions, as there is no evidence the low-frequency noise observed is due to TLS}\mrep{.  The observed low-frequency noise is likely due to electronics noise, so the TLS noise may in fact be lower and be consistent with the scaling from~\cite{defrance}.  Future measurements using correlated noise removal with the two-tone IQ mixer setup we introduced here may provide a more stringent test}.

We then scaled both our upper limits from the dark data as well as the results of \cite{defrance} to operating conditions under optical load and compared to \gsout{GR+}photon \mrep{and GR }noise measured at $T_\text{Load}\ \gsoutmath{=}\mrep{\approx}\ 180$~K.  We conservatively concluded from this comparison that these detectors should be dominated by \gsout{GR and }photon \mrep{and GR }noise down to tenths of a Hz and possibly lower, corresponding to between 1$^\circ$ and 5$^\circ$ on the sky at 0.5$^\circ$/s scan speed.  These results bode well for the use of our Al/a-Si:H MS-PPC-LEKID\gsout{s }\mrep{\ design }for astronomical observations in NEW-MUSIC, and they more generally validate the utility of a-Si:H PPCs for KIDs.

\section*{Acknowledgments}
This work was supported by funding from the National Aeronautics and Space Administration (NASA), the NASA Jet Propulsion Laboratory (JPL), the Department of Energy (DOE), the Wilf Foundation, and by Caltech. A portion of this research was carried out at the Jet Propulsion Laboratory, California Institute of Technology, under a contract with the National Aeronautics and Space Administration (80NM0018D0004). The authors acknowledge the work of numerous former students and collaborators in the development of the Al/a-Si:H MS-PPC-LEKID architecture.

\end{document}